# AN APPROACH IN OPTIMIZATION OF AD-HOC ROUTING ALGORITHMS

Sarvesh Kumar Sharma[1]

[1]Assistant Professor, Electronics and Communication Engineering, BITS-Pilani
sarveshvoice@gmail.com

## ABSTRACT

*In this paper different optimization of Ad-hoc routing algorithm is surveyed and a new method using training based optimization algorithm for reducing the complexity of routing algorithms is suggested. A binary matrix is assigned to each node in the network and gets updated after each data transfer using the protocols.*

*The use of optimization algorithm in routing algorithm can reduce the complexity of routing to the least amount possible.*

## 1. INTRODUCTION

Ad-hoc networks are the emerging area in wireless mobile networks eliminating the complexities of infrastructure and administrative modules. This technology is rapidly experiencing the real world implementation and one of the leading researcher's areas. Although it has its challenges of device heterogeneity, mobility, random traffic profiles and power conservation.

### 1.1. Terminology

Ad-hoc networks are a new paradigm of wireless communication for mobile hosts. Fixed base station is no more requirement of the wireless network as base station in mobile switching. Each user communicates directly via wireless links between them and transfer messages to next user spaced at far distance. Node mobility causes frequent changes in topology.

Ad hoc network is a collection of wireless nodes that can dynamically be set up anywhere and anytime without using any pre-existing network infrastructure [1, 2, 3, 4, 5]. Ad-hoc network is mainly characterized by its dynamic topologies, bandwidth constraint, power consumptions and security at physical layer [6, 7].

### 1.2. Routing Protocols

The Ad-hoc routing protocols are classified into four main groups: 1. Proactive or Table-driven, 2. Reactive or On-demand, 3. Hybrid, 4. Hierarchical or Cluster-based routing protocols.

Some of the most common routing protocols are named and addressed bellow:

The Destination Sequenced Distance-Vector routing protocol (DSDV) is one of the first protocol proposed in Ad-Hoc network. It is an advanced version of Bellman-Ford algorithm [10], in which each node maintains a table of shortest distance and path of first node to every other node in the network [1]. DSDV is a proactive protocol. The other type of proactive protocol is Cluster-head Gateway Switch Routing protocol (CGSR), in which nodes are organized into clusters and the coordination among members are maintained by assigning a cluster-head. There is another table-driven routing protocol called as Optimized Link State Routing protocol (OLSR), which is an advanced version of Dijkstra Algorithm [10]. The routing is based on the information existing at the table of the nodes [11, 12]. Another type of table-driven protocol, in which a set of tables are used to maintain more accurate information of the routing [13], is termed as Wireless Routing Protocol (WRP).

The on-demand driven routing protocols finds path by exchanging the routing information only when a node requires a path to communicate with the destination. The Dynamic Source Routing Protocol (DSR) [1, 2, 4, 12, 14, 15, 16, 17], Ad-hoc On-Demand Distance Vector routing protocol (AODV) [4, 12, 14, 16, 17, 18] and Temporally-Ordered routing protocol (TORA) [14, 16] are the examples of on-demand or reactive routing protocols.

The Hybrid routing protocol is based on maintaining the network topology information within the zone of the node, i.e. up to m hops. The source node and the destination node must fall in the zone. The Zone Routing protocol (ZRP) is the example of hybrid routing protocol [8].

The routing protocol is also classified based on the topology and mean of updating the routing information, as in Hierarchical routing protocol. This protocol integrates routing and QOS supports. The Core Extraction Distributed Ad Hoc routing protocol (CEDAR) [20] and Landmark routing protocol (LANMAR) [21] are the examples of such hierarchical routing protocol.

### 1.3. The Routing Protocol Problems

The existence of several routing protocols with advantages and disadvantages over each other makes network more complicated as each routing protocol algorithm requires special infrastructure.

From the study of protocol algorithms, some of them are optimized using certain methods, theories and algorithms. The main problem faced by most of these protocols optimizing algorithms is that one cannot use same protocols for all conditions and interchange between the protocols does not exist.

### 1.4. The proposed Solution

In this paper, a new methodology for optimization of routing protocols by training is suggested. After a certain period of time, the routing protocol complexity reduces to most optimizing range. Although, repetition of routing protocol for n times is not acceptable. In the respect of time consumption of the suggested method, the optimization is also repeated n times make it an overhead itself. The proposed solution presents a method for optimized repetition of routing protocol.

The presented algorithm is based on theoretical approach only, but it has promised a lot for the Ad hoc networking to be accepted and finally ready to test.

### 1.5. The Proposed Solution Claim

I agree with the solution does exist. Each node is assigned a binary matrix in the network. The route is obtained by running a routing protocol and finally updating the matrices with the assumed given route, which satisfies the need of routing protocol repetition. Refer to [9] for more information on rerouting protocols.

### 1.6 The Objective of the Paper

The main objective of the paper is to achieve a new methodology for optimized complexity in routing algorithms. Using this method, the nodes of the network do not ask for the route and send their data from their learning using the algorithm. Most important aspect is that the nodes on the route are only visited and the data is passed via nodes.

### 1.7. The Outline of the Paper

The paper is organized in three sections. In first section the problems with a proposed solution are stated. In the second section the training algorithm and its advantages are described. In the third section conclusion of the suggested algorithm in the ad hoc networks is given.

## 2. TRAINING ALGORITHM

### 2.1. Algorithm Representation

All the routing algorithms mostly optimize the obtained route and try to ignore the complexity in routing. We need to use protocol every time with every data transfer takes place. There is an optimization done in DSR [22], but it suffers from its specific protocol requirements. A cache is used to save the routed information for the repeated route.

In this paper we have targeted a protocol free algorithm. We can say it as high level, because of its availability in all routing protocols. So, one can use several routing protocols in a single network. We train the network with a routing protocol for few times between the nodes, thereafter the nodes themselves know the routing of the data transferred. And this is the time when the complexity of the routing protocol reduces to "L" equivalent to the exact route distance between the source and the destination. Hence, it is the most optimized complexity a programmer may seek for in the experiment of wireless networks.

The cost of the network is also considered in this algorithm, we have used binary matrix in training, which reduces the memory cost to minimum and increase the iteration rates.

Before writing the algorithm, the suppositions at the network are mentioned:

1- A graph of nodes and edges are used to represent the network.

2- Nodes are represented by numbers.

3- Each node is assigned by a number and that helps in ordering the edges.

4- We have assigned an array of training to each node with an array of controlling bandwidth and traffic. At the time of training, the bandwidth array is not used. The array of controlling bandwidth is used for increasing the transfer quality.

5- An adjacent matrix is generated for whole graph.

6- The nomenclature is used in the test are "G"-graph, "w"-weight, "s"-source, "d"-destination, "R"-matrix for routing and "Ro"-route array.

Here is the code of Training Algorithm:

Training_Algorithm (G,w,s,d)

If ! R[d]

Ro = get_route_from_a_routing_algorithm(G,w,s,d)

send_update(G,Ro,0,d)

-----------------------------------

Send_data(data,s,d)

Training_ Algorithm (G,w,s,d)

Transfer_data(data,s,d)

-----------------------------------

Transfer_data(data,s,d)

If s <> d

J=index_vertex(s, lg R[d])

Transfer(data,j,d)

-----------------------------------

Send_update(G,Ro,start,d)

If Ro[start] <> d

K=index_Edge(Ro[start], Ro[start+1])

R[d]=2^k

Send_update(G,Ro,start+1,d)

-----------------------------------

Index_Edge(i,j) As integer

K=0

Flag = false

While (flag)

Do

If A[i][k]

Counter=counter+1

K=k+1

If (counter=j)

Flag=false

Return counter

-----------------------------------

Index_vertex(i,k) As integer

K=0

Flag = false

While (flag)

Do

If A[i][k]

Counter=counter+1

L=L+1

If (counter=k)

Flag=false

Return L

This algorithm identifies the network being trained or not, if the network is trained, then data transfer starts, else the route of the routing protocols saved in an array. It starts training the nodes once again and send the update through the route given.

Each node has a binary matrix with one dimension. The rows of that matrix or the indexes of the array show the numeric label of each node. The bits given in rows show the edge that the node can send its data through it. By using logarithm and powering in the algorithm, the binary matrix can be filled so easily. In addition, using the binary matrix takes less memory cost and the access takes O(1) times and this causes the training and routing operations run simply.

Figure1 provides an example.

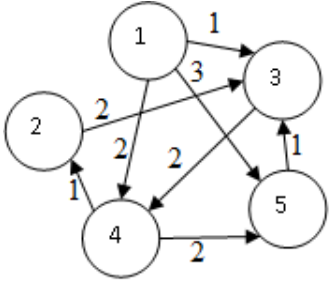

Figure 1 Graph

$$\begin{matrix} 1 \\ 2 \\ 3 \\ 4 \\ 5 \end{matrix} \begin{bmatrix} 000 \\ 001 \\ 000 \\ 001 \\ 010 \end{bmatrix}$$

| 1 | 2 | 3 | 4 | 5 |
|---|---|---|---|---|
| 0 | 2 | 1 | 2 | 4 |

$$\begin{matrix} 1 \\ 2 \\ 3 \\ 4 \\ 5 \end{matrix} \begin{bmatrix} 000 \\ 001 \\ 000 \\ 000 \\ 000 \end{bmatrix}$$

| 1 | 2 | 3 | 4 | 5 |
|---|---|---|---|---|
| 0 | 0 | 1 | 1 | 1 |

$$\begin{matrix} 1 \\ 2 \\ 3 \\ 4 \\ 5 \end{matrix} \begin{bmatrix} 000 \\ 000 \\ 000 \\ 000 \\ 000 \end{bmatrix}$$

| 1 | 2 | 3 | 4 | 5 |
|---|---|---|---|---|
| 0 | 1 | 0 | 1 | 1 |

International Journal of Distributed and Parallel Systems (IJDPS) Vol.3, No.3, May 2012

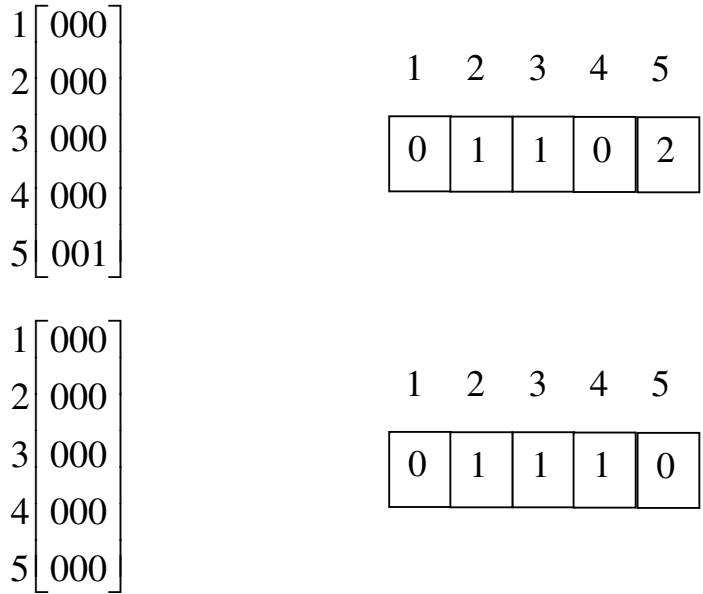

Fig.1 A sample for the matrices designed for the arrays of the nodes in a graph G.

While sending update, the matrix in each node on the route is updated. For example if the route from 1 to 2 first passes from 4, and the edge from 1 to 4 is the second edge of 1, then in the matrix at the fifth row the number $[9]_2$ (=$[2]10$) should be written; which means from 1 to 2 we should pass the second edge.

The function index_edge gets the two adjacent nodes and gives the number of edge between them which is ordered by the number of the nodes. For instance in order to find out that the second edge of 1 is related to which node, we refer to the adjacent matrix.

Figure2 provides the adjacent matrix for the graph above.

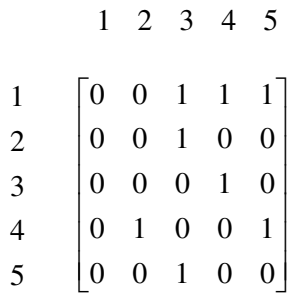

Figure 2. A sample for the adjacent matrix designed for the graph G.

In the adjacent matrix the algorithm counts the number of 1s and the number of the comparing operation repetition. At the matrix in the example when it reaches to the second 1, it has done the comparing operation for 4 times. So it recognizes that the second edge of node 1 is related to the node 4. Therefore in order to reach to the node 2, it should pass the node 4. The function index_vertex gets the number of a node and an edge connected to it and gives the number of the adjacent node related to that edge. To find the order of the edge, this algorithm uses logarithm in the basis of 2. In the example above (log 100) equals with 3 which means to the mentioned node we should pass the third edge. By this method we avoid the overload of clustering and the high complexity of for-loop.

We have also an adjacent matrix for the whole graph which its memory cost in a graph with so many nodes would be inconsiderable though its processing would be time consuming.

While the data transferring is active and one of the nodes runs away from the network, the number of rows in training matrix in which the route passes from that node (the number of the edge related to that node is 1) sets to zero. So the routing and training algorithms for the related destinations should be repeated.

## 2.2. Benefits of Using This Method

As it is mentioned, this algorithm doesn't optimize the routing protocol, but it guides and trains the network to learn the protocol once and use the route given for thousands of times. In addition while sending update amongst the route found, the nodes on the route are also updated. The important point is that if the destination is far from the source and it passes through the most of the nodes, then large amount of nodes become updated and this causes less repetition of routing protocol.

Being protocol-free, this method can be theoretically applied in all kinds of mobile or wireless ad-hoc networks and there would be no limitation in using this algorithm.

Propagating packets or messages for requesting and replying through the network causes the traffic to be heavier.

But in our method after initial routing we never use such packets or messages. As it is said before, the time complexity reduces to "L" after training, which means that the data has been transferred through the right path. Using the binary matrix for each node, the memory complexity for n nodes is n bytes. Less memory occupied and less time spent, makes this method a desirable way in data transfer on networks).

## 3. CONCLUSION

Repeating the routing protocols for each node is a time consuming task. Currently applicable algorithms, in this regard, suffer from a high complexity of time which is discussed in section 2. In section 3 we have introduced a new algorithm, which by training the network reduces the number of times that the routing protocols are being repeated. After a few repetition of routing the nodes themselves learn from where they should send data.

## 3.1. Future Work

The method we presented in this paper sets the stage ready for an interesting topic of research:

1. Traffic Control on Ad-Hoc Networks.

2. A New Approach:

The algorithm introduced in this paper can set the bandwidth array, in addition to the training. The bandwidth array is adjusted to control the traffic of the route. While sending data and routing to the destination the algorithm increments a counter. This means that the path from the source to the destination has been used for the number of times the counter shows. Now, if we notice that one path is used so many times and one less, we can maintain the bandwidth of the path. This is a good way to control the traffic on the network. If the bandwidth has reached to the highest amount, new edges may be produced. It seems to be a good idea to manage the traffic on the network.

**Authors**

Mr. Sarvesh Kumar Sharma (Assistant Professor, ECE), received the Master's of Engineering degree in Communication Engineering from BITS-Pilani, India in 2009, BE degree in ECE in 2006. Nowadays his research activity is focused on the Wireless Sensors Network protocols and optimization techniques in the real environment. He is also active researcher in e-Governance using ICT for technical education in India. Short Biography

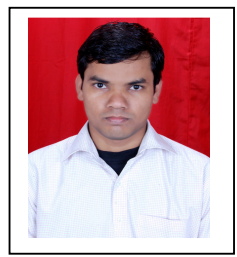